\documentclass[twocolumn,showpacs,preprintnumbers,amsmath,amssymb,pra]{revtex4}

\usepackage{epsfig}
\usepackage{graphics}
\usepackage{graphicx}

\usepackage{dcolumn}
\usepackage{bm}

\newcommand{\be}{\begin{equation}}
\newcommand{\ee}{\end{equation}}
\newcommand{\bea}{\begin{eqnarray}}
\newcommand{\eea}{\end{eqnarray}}

\begin{document}
\newcolumntype{d}[1]{D{.}{\cdot}{#1}}
\newcolumntype{.}{D{.}{.}{-1}}
\newcolumntype{,}{D{,}{,}{2}}

\title{
Ultrahigh finesse Fabry-Perot superconducting resonator}
\author{S.~Kuhr}
\author{S. Gleyzes}
\author{C. Guerlin}
\author{J. Bernu}
\author{U. B. Hoff}
\author{S.~Del\'{e}glise}
\author{S. Osnaghi}
\author{M. Brune}
\author{J.-M.~Raimond}
\affiliation{Laboratoire Kastler Brossel, D\'epartement de
Physique de l'Ecole Normale Sup\'erieure,
24 rue Lhomond, F-75231 Paris Cedex 05, France}
\author{S. Haroche}
\affiliation{Coll\`ege de France, 11 Place Marcelin Berthelot, F-75231 Paris
Cedex 05, France}
\author{E. Jacques}
\author{P. Bosland}
\author{B. Visentin}
\affiliation{DAPNIA, Orme des Merisiers, CEA, F-91191 Gif-sur-Yvette
Cedex, France}

\date{\today}

\begin{abstract}
We have built a microwave Fabry-Perot resonator made of
diamond-machined copper mirrors coated with superconducting
niobium. Its damping time ($T_{\rm c}= 130$~ms at 51~GHz and
0.8~K) corresponds to a finesse of $4.6\times10^9$, the highest
ever reached for a Fabry-Perot in any frequency range.
This result opens many perspectives for quantum information processing, decoherence and
non-locality studies.
\end{abstract}


\maketitle

Since Bohr-Einstein's photon box thought experiment,
storing a photon for a long time has
been a dream of physicists. Cavity quantum electrodynamics (CQED)
in the microwave domain comes closest to this
goal. Photons are trapped in a superconducting cavity and probed
by atoms crossing the field one at a time. Experiments with
circular Rydberg atoms and Fabry-Perot resonators have led to
fundamental tests of quantum theory and various demonstrations of
quantum information procedures \cite{rmp}. The open geometry of
the cavity is essential to allow a perturbation-free propagation
of long-lived atomic coherences through the mode. With this
cavity structure, however, the field energy damping time $T_{\rm
c}$ is very sensitive to geometrical mirror defects, limiting  $T_{\rm c}$ to $\simeq 1$~ms in previous experiments.
We report here the realization of a Fabry-Perot
resonator at $\omega/2\pi=51$~GHz, with
$T_{\rm c}=130$~ms. The
cavity quality factor $Q$ is $4.2\times10^{10}$ and its finesse
$4.6\times10^9$, the highest ever achieved in any
frequency domain for this geometry. This important step opens the way to many CQED experiments. Quantum non-demolition detection of a single photon \cite{Gleyzes06} and generation of mesoscopic non-local quantum superpositions \cite{Milman2005} are now accessible. Long term storage of single photon fields opens bright perspectives for quantum information processing. These high-$Q$ cavities are also promising for the stabilization of microwave oscillators or for the search of exotic particles \cite{rmpaxion}.

\begin{figure}[!b]
\begin{center}
\includegraphics[width=0.3\textwidth]{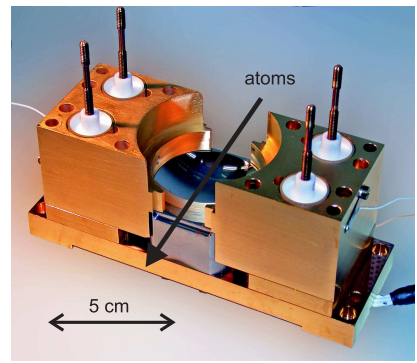}
\end{center}
\vspace{-0.3cm}
 \caption{\label{fig:setup} Photograph of the cavity
assembly with the top mirror removed. The atomic beam path is
visualized by the arrow. The four posts are used to mount the
upper mirror. The piezoelectric actuators,
centered by the white Teflon cylinders, surround the posts.}
\end{figure}

A picture of the cavity $C$ with the top mirror removed is shown in
Fig.~\ref{fig:setup}. The mirrors have a diameter $D_0=50$~mm. The distance between their apexes is $L=27.57$~mm. Their surface is toroidal (radii of curvature 39.4 and 40.6~mm in two orthogonal planes).
The two  TEM$_{900}$ modes near $51.099$~GHz with orthogonal linear
polarizations are separated by $1.2$~MHz. This large frequency splitting is essential to ensure that atoms are efficiently coupled to a single mode only.
The mirrors are
electrically insulated. A static electric field parallel to the
cavity axis is applied between them to preserve the circular
states and to tune the atomic transition via the Stark effect
\cite{rmp}. The 1~cm spacing between mirror edges is partly closed
by two guard rings improving the static field
homogeneity in $C$. The atoms of a thermal beam enter and exit
the cavity through two large ports ($1$~cm$~\times~2$~cm) so that
they never come close to metallic surfaces, preserving them from
patch effect stray fields. This ensures a good transmission of
atomic coherences through the cavity \cite{Gleyzes06}. Four piezoelectric actuators
are employed to translate one of the mirrors and to tune the
cavity (within $\pm 5$~MHz) with a few Hz accuracy.

We realized two mirror sets, $M_1$ and $M_2$. The copper substrates
are first machined to a gross spherical shape. They are then
submitted to two temperature cycles to release stresses. They are
first heated in a vacuum chamber to $400^{\circ}$C and cooled down
to liquid $N_2$ temperature. The final diamond machining is then
performed (Kugler company). The local surface roughness is $10$~nm
r.m.s. and the surface has a peak-to-valley deviation of $< 300$~nm from the
ideal shape. To avoid deformation of the mirrors in the
final assembly, their thickness is $30$~mm. Their back surface and
that of the matching holders are polished to $1\ \mu$m.

The mirror surface is covered with a $12~\mu$m-thick layer of Nb. We use
a coating facility at CEA, Saclay, designed for r.f.
cavities used in particle accelerators \cite{CEA_accelerators}. The Nb
layer is deposited by d.c. cathode sputtering in a magnetron discharge
\cite{CEA_sputtering}. We first clean the substrate with
ultra-pure filtered alcohol (ultrasonic bath) and dry it with
filtered Ar. The sputtering chamber is evacuated to $10^{-8}$~mbar
and then filled with $10^{-1}$~mbar of Ar. We set the mirror
potential to $-1000$~V for $20$~s, creating an Ar plasma which
blows away residual dust particles. The chamber is evacuated
again and we start the sputtering process. A $1$~kW magnetron
creates a dense Ar plasma in the vicinity of a cylindrical Nb
cathode (Ar pressure during sputtering: 10$^{-3}$ mbar).
Initially, the Nb cathode is far away from the
mirror. When its impure surface layer is removed, we move it
in front of the mirror. The evaporated atoms are deposited
at a rate of $0.1~\mu$m/min on the mirror surface, which heats up
to $300-400^{\circ}$C. Before being mounted in
the Rydberg atom set-up, the mirrors are finally rinsed in an alcohol
ultrasonic bath and dried with Ar.

In order to characterize the cavity modes,
microwave is coupled in via weak diffraction
loss channels. We thus avoid coupling irises in the mirror centers,
which are detrimental to the surface quality  \cite{rmp}.
 This coupling is large enough to
inject a mesoscopic field in $C$, but not
to detect directly the
decay of the leaking field. We monitor instead
the cavity ring-down with an atomic probe.  At the beginning of a
measurement sequence, we inject a microwave pulse
by a wave\-guide ending in the guard ring around $C$.
Most of the microwave power is not
coupled into the mode and decays in the apparatus
in less than 1~$\mu$s.
After a time interval $t$, we send the Rydberg atom
probe in $C$. It is produced by a two-step laser
excitation of $^{85}$Rb atoms  involving a diode laser at
$420.30$~nm ($5S_{1/2}$ $\rightarrow$ $6P_{3/2}$)
and a second diode laser at $1014.67$~nm
($6P_{3/2}$ $\rightarrow$ $52D_{5/2}$). The
cavity field induces transitions from $52D_{5/2}$
to other Rydberg levels. The absorption is made
broadband by the Stark effect in a 13.4~V/cm
electric field applied in $C$. Broadband detection
is essential for the first cavity tests, since the
reproducibility of the mirror mounting results in
an uncertainty in the mode frequencies of $\pm
10$~MHz.
The cavity-field-induced atomic transitions are monitored by a
state-selective field-ionization detector.

By sweeping the microwave source and recording the
atomic absorption, we determine the cavity
resonances. For $M_1$ and $M_2$ we find two modes
(lower frequency, $M_i^{LF}$ and higher frequency, $M_i^{HF}$, $i=1,2$) close to
51.1~GHz, separated by the expected 1.2~MHz splitting. We
have mapped the transverse profile of these modes. Using pulsed
velocity-selected atomic samples, we know the atomic position at
any time. Switching the Stark field in $C$ on and off, we set the
atoms in resonance during a short time window, at a well defined
position in $C$. The measured transition probability
reveals the intensity profile of
the mode along the atomic beam axis. We obtain, as expected,
a Gaussian with a
$w_0\simeq6$~mm waist.

\begin{figure}[bt]
\begin{center}
\includegraphics[width=0.4\textwidth]{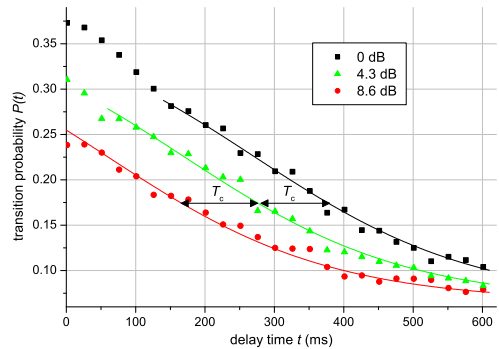}
\vspace{-0.5cm}
\end{center}
\caption{\label{fig:curves_s} Measurement of $T_{\rm c}$
($M_1^{LF}$) at $T=0.8$~K. Transition probability $P(t)$ as a function of the delay $t$
between microwave injection and the atomic probe crossing the
cavity. The points (circles, triangles and squares) correspond to
three initial field energies $E_0$, $eE_0$ and $e^2 E_0$,
respectively. The fits (solid lines) result from a simple
absorption model including saturation. They are equidistantly
shifted in time with respect to one another by $T_{\rm c}=112\pm4$~ms. Each point is the average of 1600 atomic detections. }
\end{figure}

To measure the quality factors, we record the
transition probability $P(t)$ as a function of the delay $t$.
Typical data, obtained with $M_1^{LF}$
at $0.8$~K is shown in Fig.~\ref{fig:curves_s} for three
initial field energies, $E_0$, $eE_0$ and $e^2 E_0$ ($e$
is the base of natural logarithms), corresponding to microwave
source attenuations of 8.6, 4.3 and 0 dB respectively. Due to the
exponential decrease of $E$ versus time, $E(t)=E_0e^{-t/T_{\rm
c}}$, an arbitrary energy $E<E_0$ is reached at times $t_0$,
$t_0+T_{\rm c}$ and $t_0+2T_{\rm c}$ for these three attenuations.
Since $P$ depends only upon $E$, the corresponding $P(t)$ curves
are time-shifted by $T_{\rm c}$ with respect to each
other. We obtain $T_{\rm
c}=112\pm 4$~ms. For $M_1^{HF}$, $T_{\rm
c}$ is found to be $87\pm10$ ms. For $M_2^{LF}$ and  $M_2^{HF}$
we get, at $0.8$~K, $T_{\rm c} =74\pm
6$~ms and $T_{\rm c} =130\pm 4$~ms respectively. These four modes have all an
extremely long energy storage time. The longest one corresponds to a light
travel distance of 39\,000 km folded in the 2.7~cm space between the mirrors.
The corresponding
quality factor is $Q=\omega T_{\rm c}=4.2\times 10^{10}$ and the
finesse is $f=Q/9=4.6\times 10^9$.

We have measured the $M_2^{HF}$ mode spectrum at 0.8~K. The FWHM linewidth is $3\pm 0.5$ Hz, close to the 1.22 Hz value deduced from $T_c$. The difference is due to residual low frequency mechanical vibrations (a 1 Hz shift corresponds to a 500 fm translation of one mirror). The cavity drift is less than 3 Hz per hour. The stored field coherence is thus well preserved, an important feature for quantum information storage.

\begin{figure}
\begin{center}
\includegraphics[width=0.4\textwidth]{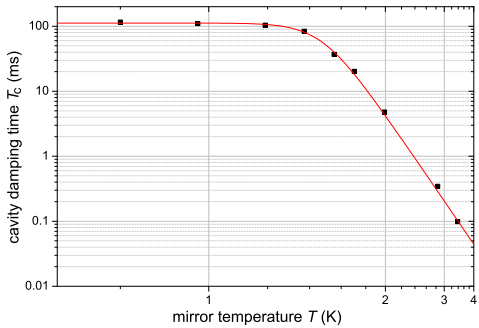}
\vspace{-0.7cm}
\end{center}
\caption{\label{fig:TCav_vs_Temp} Cavity damping time $T_{\rm c}$
versus mirror temperature $T$ ($M_1^{LF}$). The horizontal scale is
reciprocal and the vertical one
logarithmic. The dots are experimental. The solid line is a
fit  providing a gap value $\Delta_0/k_{\rm
B}=20.2$~K. Below 1.4~K, $T_{\rm c}$ saturates at 112~ms.
 }
\end{figure}

We have studied $T_{\rm c}$ as a function of the mirror temperature $T$
for $M_1^{LF}$ (Fig.~\ref{fig:TCav_vs_Temp}).
For $T>1.4$~K, $T_c$ increases exponentially versus $1/T$, while it saturates for $T<1.4$~K.
The quality factor $Q$ can be expressed as the ratio
$G/R_{\rm e}$ of two resistances characterizing the geometry ($G=2800$~$\Omega$ here)
and the losses ($R_{\rm e}$) \cite{Halbritter1971}. The effective resistance, $R_{\rm
e} = R_{\rm BCS}+R_0+R_d$ has three contributions.
The resistance $R_{\rm BCS}$ is given by BCS theory, the residual resistance $R_0$
at $T=0$~K is due to defects and $R_d$ measures the
diffraction losses. The BCS resistance is $R_{\rm
BCS}=(A/T)\exp(-\Delta_0/k_{\rm B}T)$, where $\Delta_0$ is the
superconducting gap and $A$ is intrinsic to Nb. For $T > 1.6$~K, $R_{\rm BCS}$ dominates $R_{\rm e}$, while it is negligible for $T<1.4$~K.
The line in Fig.~\ref{fig:TCav_vs_Temp} is a fit with
$T_{\rm c}=G/\omega R_e$, from which we
infer $\Delta_0/k_{\rm B}= 20.2\pm0.3$~K. This value, confirmed by
measurements on $M_1^{HF}$ and $M_2$, differs slightly from the value
17.056~K found at 22~GHz with bulk Nb in Ref.~\cite{Meschede1985}.

The saturation of $T_{\rm c}$ below $1.4$~K yields
$R_0+R_d = 75$~n$\Omega$ for $M_1^{LF}$
and $68$~n$\Omega$ for $M_2^{HF}$. The order of magnitude of the
diffraction losses $R_d$ can be estimated. The mirror diameter
$D_0$ limits $Q$ to $Q_{\rm diff} = (\omega L/c)
\exp{(D_0^2/2w^2)} = 2.7\times10^{11}$ ($w=1.23\,w_0$ is the mode
waist at the mirror surface). The surface roughness is
characterized by the r.m.s. deviation, $h_{\rm rms}$, with respect to the ideal shape. The corresponding quality
factor, calculated by evaluating the ``total integrated
scattering'' (TIS) \cite{Winkler1994}, is $Q_{\rm surf} =
cL/4\omega h_{\rm rms}^2$. From the measured $h_{\rm rms} \simeq 10$~nm,
we obtain $Q_{\rm surf} =
6.4\times 10^{10}$. Combining these losses yields
$Q'=(Q_{\rm diff}^{-1} + Q_{\rm surf}^{-1})^{-1} =
5.2\times10^{10}$. This is close to the best measured value
($Q=4.2\times 10^{10}$), indicating that
diffraction losses are the dominant contribution at low $T$.

It is also instructive to compare the $Q$ factor of our open
resonators with that of closed cylindrical cavities at $22$~GHz
used in Rydberg atom micromaser studies
\cite{Meschede1985,Rempe1990}. In these experiments,
$Q=4.0\times10^{10}$ at $0.3$~K was measured
\cite{Walther2001}. The residual resistance
was $R_0 = G/Q = 28$~n$\Omega$ ($G= 1089$~$\Omega$ -- there are no
diffraction losses in this geometry), a value comparable to ours.
The difference could be due to
diffraction losses in our open geometry,
to the frequency
dependence of $R_0$ or to variations in the Nb purity.
It is remarkable that our open
cavity reaches the same $Q$ as the best closed one in the same
frequency domain. At much lower frequencies, around 1~GHz, $Q$
factors up to $10^{12}$ have been obtained \cite{Cavities1GHz}.
This frequency domain is however much less convenient for cavity
QED experiments.

We have reported the realization of a ultra-high-$Q$ photon box.
This cavity, with its open geometry, is ideally
suited for the propagation of atomic coherence through the field
mode, atomic interferometry, decoherence studies and quantum information
processing experiments. Experiments with two such cavities are of
particular interest. A single atom could be used to entangle two mesoscopic
fields separated by a macroscopic distance, resulting in the
preparation of a non-local quantum state \cite{Milman2005}.

Laboratoire Kastler Brossel is a laboratory of Universit\'{e}
Pierre et Marie Curie and ENS, associated to CNRS (UMR 8552).
We acknowledge support by the Japan Science and Technology Agency
(JST), by the EU under the IP projects ``QGATES'' and ``SCALA'',
and by a Marie-Curie fellowship of the European Community (S.K.).

\end{document}